\documentclass[aps,pra,superscriptaddress,showpacs,tightenlines,cap,cs5size,winfonts,nospace,indent,fancyhdr,twocolumn]{revtex4-2}

\usepackage{amsmath}
\usepackage{changepage}
\usepackage{epsfig}
\usepackage{epstopdf}  
\usepackage[colorlinks,citecolor=blue,linkcolor=blue]{hyperref}
\usepackage{lipsum}
\usepackage{amssymb}
\usepackage{mathtools}
\usepackage{subfigure}
\usepackage{multirow}
\usepackage[table]{xcolor}
\usepackage{color}
\usepackage{graphicx}
\begin{document}

\title{Nontraditional Deterministic Remote State Preparation Using a Non-Maximally Entangled Channel without Additional Quantum Resources}
\author{Xuanxuan Xin}
\author{Shiwen He}
\author{Yongxing Li}
\author{Chong Li}
\thanks{lichong@dlut.edu.cn}
\affiliation{School of Physics, Dalian University of Technology, Dalian 116024,China}

\begin{abstract}
    In this paper, we have reinvestigated probabilistic quantum communication protocols and developed a nontraditional remote state preparation protocol that allows for deterministically transferring information encoded in quantum states using a non-maximally entangled channel. With an auxiliary particle and a simple measurement method, the success probability of preparing a d-dimensional quantum state is increased to 1 without spending additional quantum resources in advance to improve quantum channels, such as entanglement purification. Furthermore, we have designed a feasible experimental scheme to demonstrate the deterministic paradigm of transporting a polarization-encoded photon from one location to another using a generalized entangled state. This approach provides a practical method to address decoherence and environmental noises in actual quantum communication.
\end{abstract}

\maketitle

\section{Introduction}
The quantum network, the backbone of quantum computing architectures, is composed of spatially separated nodes storing quantum information in quantum bits and connected by quantum channels \cite{PhysRevA.29.1419, PhysRevLett.78.3221, ritter2012elementary, simon2017towards, PhysRevLett.120.030501, Wei2022}. Quantum communication tasks transferring information between disparate quantum nodes constitute the essential elements necessary for integrated quantum networks \cite{PhysRevLett.67.661, duan2001long, Quantumdirectportation, ursin2007entanglement}, such as quantum teleportation (QT) \cite{PhysRevLett.70.1895, PhysRevA.58.4394, PhysRevA.58.4373}, quantum key distribution (QKD) \cite{PhysRevLett.85.5635, PhysRevLett.85.441, PhysRevLett.94.230504, PhysRevLett.95.010503, Qi_2021, Kwek2021}, quantum dense coding (QDC) \cite{PhysRevLett.69.2881, PhysRevLett.76.4656, PhysRevLett.93.210501, Guo} and quantum secure direct communication (QSDC) \cite{PhysRevA.65.032302, PhysRevA.68.042317, PhysRevA.69.052319, PhysRevLett.118.220501}. Suppose two communicators named Alice and Bob are placed in two different quantum nodes and previously share an entanglement in the form of the quantum channel. In these ideal quantum communication tasks, maximally entangled states are employed as quantum channels, which guarantees that schemes are deterministic. However, the actual entangled channel inevitably degenerates to an undesired non-maximally entangled channel due to decoherence and environmental noise \cite{PhysRevA.52.R2493, PhysRevLett.70.1187, PhysRevLett.81.2594, RevModPhys.76.1267, PhysRevA.72.012315}. The corresponding consequence is the success probability of quantum communication degraded from 100\% to a lower value. The information may be transferred unsuccessfully and lost \cite{PhysRevA.62.024301, PhysRevA.61.034301,PhysRevA.68.022310}.

Entanglement is indispensable in scalable quantum calculation and quantum communication assignments \cite{kimble2008quantum, e17041755, pirandola2016physics, wehner2018quantum, e25010061, Lu2023, Xu2022}. It is simple to produce with the current experimental techniques, whereas it is challenging to entangle different long-distance quantum systems strongly \cite{dakic2012quantum, PhysRevLett.106.130506,yao2012observation,huang2011experimental, Zhang2023}. In other words, the ideal maximally entangled channels are challenging to prepare in practice. Even if prepared successfully, it would decay to a non-maximum entanglement due to the noise and decoherence \cite{wagenknecht2010experimental,zhang2011preparation,yin2020entanglement}. When the ideal maximally entangled channel is replaced with a non-maximally entangled channel, the efficiency of conventional quantum communication is reduced, such as in probabilistic teleportation \cite{PhysRevA.62.024301, PhysRevA.61.034301,PhysRevA.68.022310}. The mainstream solution currently is to increase the success probability of communication by improving entangled channels. There exist sorts of practices to enhance entanglement, such as entanglement purification \cite{PhysRevLett.76.722,PhysRevLett.77.2818,PhysRevA.54.3824,PhysRevLett.97.180501,PhysRevLett.110.260503,PhysRevLett.126.010503}, quantum catalysis \cite{jonathan1999entanglement,PhysRevA.64.042314,PhysRevA.67.060302,PhysRevA.79.054302} and local filtering operations \cite{PhysRevLett.74.2619,PhysRevA.54.2685,PhysRevLett.96.150501,PhysRevLett.100.090403,PhysRevA.86.052115}. Entanglement purification protocols achieve the above objective by locally manipulating multiple copies of harsh entangled states to produce fewer copies with increased fidelity \cite{PhysRevLett.127.040502}. The method with the advantage of quantum catalysts has high communication efficiency without ancillary entanglement being consumed or degraded \cite{PhysRevLett.127.080502}. However, more entanglement resources are requisite beforehand in both approaches, which increases the difficulty of experiments. Local filtering operations on entangled quantum channels may activate their desirable features but additionally have higher experimental complexity \cite{PhysRevResearch.3.023045}. Briefly, boosting the entangled channels requires tedious operations and more consumption of quantum resources. Therefore, fresh methods instead of improved entanglement should be considered to increase the probability of successful communication through non-maximally entangled channels with limited resource consumption.

Remote state preparation (RSP) is a useful quantum communication task to securely transfer messages encoded in quantum states between distant places without physically sending the states themselves \cite{pati2000minimum,PhysRevA.62.012313, PhysRevLett.87.077902}. This protocol is a variant of QT in which the sender knows the quantum information prepared for the receiver. The original RSP could only prepare real number information deterministically. Recently, various protocols have sprung up to prepare general complex number information perfectly \cite{nguyen2011remote, dakic2012quantum, e18070267, PhysRevA.98.042329, Du2019}. These deterministic programs address the potential failure of delivering general information in the initial scheme. However, the ideal maximally entangled channels are employed in these schemes. If these ideal channels are replaced with more general non-maximally entangled states, the above RSP schemes are changed from deterministic to probabilistic. 

To this end, we provide an alternative RSP paradigm wherein a general quantum state is prepared deterministically in a remote place via a generally entangled quantum channel. The probability of success is increased from $2|\alpha|^{2} (|\alpha|^{2}\leq\frac{1}{2})$ to 1 without increased quantum resources. It is independent of the coefficients of the entangled channel in this protocol. Even if the utilized entangled channel is degraded to a non-maximally entangled channel, the probability of success stays constant as long as this entanglement is present. No additional quantum resources are spent in our scheme to improve the entanglement channel in advance, which minimizes the expenditure of costs and lessens the experimental complexities. For specific operations, an auxiliary particle is introduced in this protocol compared to the initial particle. Additionally, the easier-to-operate projective measurement under the simple basis vectors is adopted instead of the complex positive operator-valued measurement. This deterministic RSP protocol applies to preparing quantum states of arbitrary dimensions. We first give a specific scheme for the remote preparation of two-dimensional quantum states and then extend it to the preparation of higher-dimensional quantum states. This work provides constructive implications for quantum communication technologies and quantum networks.

\section{Preliminaries}

In the ideal RSP programs \cite{PhysRevA.62.012313, PhysRevLett.87.077902, pati2000minimum}, two spatially separated communicators, the sender Alice and the receiver Bob, previously shared a maximally entangled quantum channel \mbox{$\frac{1}{\sqrt{2}}(|00\rangle+|11\rangle)_{AB}$}. Alice possesses qubit A. Bob holds the other qubit B. The goal of Alice is to transport information encoded on a qubit state to Bob using local operations and classical communication (LOCC). Suppose the prepared quantum state for Bob is a two-dimensional quantum state $|\phi \rangle=a|0\rangle+be^{i \gamma}|1 \rangle$ (where $a$ and $b$ are real numbers and $|a|^{2}+|b|^{2}=1$). Alice knows all of the parameters $a$, $b$, and $\gamma$ precisely, but Bob knows nothing. Since Alice knows $a$, $b$, and $\gamma$, she is fully able to measure her qubit on any basis determined by those parameters. Let such a basis be $\{|\mu_{0}\rangle=a|0\rangle+be^{i \gamma}|1 \rangle, |\mu_{1}\rangle=be^{-i \gamma}|0\rangle-a|1 \rangle \}$. When Alice measures the qubit A on the basis $\{|\mu_{0}\rangle, |\mu_{1}\rangle\}$, the outcomes $|\mu_{0}\rangle_{A}$ or $|\mu_{1}\rangle_{A}$ occurs at random, but they each have a probability of 50\%. According to the original scheme, Bob would only receive the message when the measurement result is $|\mu_{0}\rangle_{A}$. When the measurement result is $|\mu_{1}\rangle_{A}$, the RSP fails. Nguyen et al. improved this protocol by introducing an auxiliary qubit C initially set in the $|0\rangle_{C}$ \cite{nguyen2011remote}. In this improved program, Alice performs the C-NOT gate $C^{(2)}_{AC}$ on qubits A and C, where $C^{(2)}_{ij}=|0\rangle_{ii}\langle 0|\otimes I_{j}+|1\rangle_{ii}\langle 1|\otimes \sigma^{x}_{j}$. As a result, a hyperentangled state is established: $\frac{1}{\sqrt{2}}(|000\rangle+|111\rangle)_{ABC}$. Then, Alice measures qubit A based on the basis $\{ |\mu_{0}^{'}\rangle=a|0\rangle+b|1 \rangle, |\mu_{1}^{'}\rangle=b|0\rangle-a|1 \rangle \}$. If the measurement result is $|\mu_{0}^{'}\rangle_{A}$, Alice applies the phase gate $P_{C}=|0\rangle \langle 0| + e^{2i \gamma}|1\rangle \langle 1|$ on qubit C. If the measurement result is $|\mu_{1}^{'}\rangle_{A}$, Alice does nothing. After that, Alice measures qubit C based on the basis $\{|\nu _{0}\rangle=|0\rangle+e^{i \gamma}|1\rangle, |\nu_{1} \rangle=e^{-i \gamma}|0\rangle-|1\rangle \}$. Then she tells Bob the measurement outcomes of qubits A and C. There are four combinations of two qubit measurements, $|\mu_{0}^{'}\nu _{0}\rangle_{AC}$, $|\mu_{0}^{'}\nu _{1}\rangle_{AC}$, $|\mu_{1}^{'}\nu _{0}\rangle_{AC}$, and $|\mu_{1}^{'}\nu _{1}\rangle_{AC}$. Each of Alice's outcomes happens randomly but with an equal probability of 25\%. In either case, Bob can reconstruct the target state $|\phi \rangle=a|0\rangle+be^{i \gamma}|1 \rangle$ by applying corresponding operations to qubit B based on the measurements \cite{nguyen2011remote}. Thus, the total success probability is P = 25\% $\times$ 4 = 100\%, rendering this protocol a deterministic one.

As we know above, the utilization of maximally entangled channels enables a 100\% probability of successful communication. Regrettably, realistic quantum channels are noisy on account of decoherence and the environment. In other words, ideal maximally entangled channels are deteriorated to non-maximally entangled ones in practice. When the quantum channel is a partially entangled state $(\alpha|00\rangle+\beta|11\rangle)_{AB}$ ($|\alpha|^{2}+|\beta|^{2}=1$, $|\alpha|\leq|\beta|)$, the success probability of RSP is decayed to $2|\alpha|^{2} (|\alpha|^{2}\leq\frac{1}{2})$ in these ideal RSP schemes. The proof is as follows. According to conventional deterministic remote state preparation protocols \cite{nguyen2011remote, PhysRevA.98.042329, Du2019}, Alice introduces an auxiliary qubit C set on the state $|0\rangle_{C}$ initially and performs the C-NOT gate $C^{(2)}_{AC}$ on qubits A and C,
\begin{eqnarray}
    \begin{aligned}
        |\varphi_0\rangle^{(2)}_{ABC}&=C^{(2)}_{AC}(|\varphi\rangle^{(2)}_{AB}\otimes|0\rangle_{C})\\
        &=(\alpha|000\rangle+\beta|111\rangle)_{ABC},
    \end{aligned} 
\end{eqnarray}
where
\begin{eqnarray}
    \begin{aligned}
       C^{(2)}_{AC}=|0\rangle_{AA}\langle 0|\otimes I_{C}+|1\rangle_{AA}\langle 1|\otimes \sigma^{x}_{C}.      
    \end{aligned}
\end{eqnarray}

\begin{figure}[htbp]
    \includegraphics[width=0.4\paperwidth]{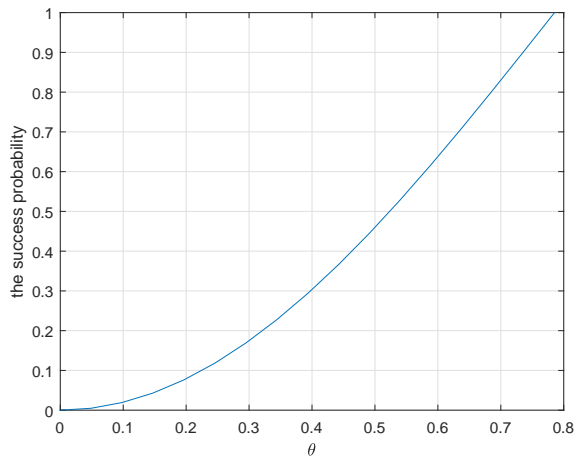}
    \caption{The success probability as a function of the entanglement coefficient $\theta$ in conventional quantum communication protocols. It is presented that the probability of success can reach 1 only when $\theta=\frac{\pi}{4}$, i.e., the quantum channel is an ideal maximally entangled quantum state ($|\alpha|=|\beta|=\frac{1}{\sqrt{2}}$).}\label{Fig.1}
\end{figure}

Then she applies a controlled-U operation $CU^{(2)}_{ij}$ on particles A and C, and subsequently, the system evolves into
\begin{eqnarray}
    \begin{aligned}
        |\varphi_1\rangle^{(2)}_{ABC}=&CU^{(2)}_{AC}|\varphi_0\rangle^{(2)}_{ABC}\\
                               =&\alpha(|000\rangle+|111\rangle)_{ABC}\\
                               &+\sqrt{\beta^{2}-\alpha^{2}}|110\rangle_{ABC},      
    \end{aligned} 
\end{eqnarray}
where
\begin{eqnarray}
    \begin{aligned}
    CU^{(2)}_{AC}=&|0 \rangle_{ii} \langle 0| \otimes I_{j}+|1 \rangle_{ii} \langle 1| \otimes \\
                  &[(\alpha/\beta|0 \rangle-\sqrt{1-\alpha^{2}/\beta^{2}}|1 \rangle)_{jj}\langle 0|\\
                  &+(\alpha/\beta|1 \rangle+\sqrt{1-\alpha^{2}/\beta^{2}}|0 \rangle)_{jj}\langle 1|],
\end{aligned} 
\end{eqnarray}
where the agreement $|\alpha| \le |\beta|$ exists. Alice then employs the C-NOT gate $C^{(2)}_{AC}$ on qubits A and C again, and the above quantum state evolves into
\begin{eqnarray}
    \begin{aligned}
        |\varphi_2\rangle^{(2)}_{ABC}=&C^{(2)}_{AC}|\varphi_1\rangle^{(2)}_{ABC}\\
                               =&\alpha|0\rangle_{C}(|00\rangle+|11\rangle)_{AB}\\
                               &+\sqrt{\beta^{2}-\alpha^{2}}|1\rangle_{C}|11\rangle_{AB}.  \label{equation A7}   
    \end{aligned} 
\end{eqnarray}
Next is the projective measurement on the auxiliary qubit C. There are two types of measurement results, $|0\rangle_{C}$ and $|1\rangle_{C}$. The probability of the measurement result $|0\rangle_{C}$ is $2\alpha^{2}$. The probability of the measurement result $|1\rangle_{C}$ is $\beta^{2}-\alpha^{2}$. If the measurement result is $|0\rangle_{C}$, the three-qubit quantum state (Equation (\ref{equation A7})) is collapsed into the maximally entangled quantum state $(|00\rangle+|11\rangle)_{AB}$, and RSP succeeds. If the measurement result is $|1\rangle_{C}$, RSP fails. Thus, RSP succeeds only when the measurement result is $|0\rangle_{C}$. The success probability of RSP is $2\alpha^{2} (\alpha^{2} \le \frac{1}{2})$ via a non-maximally entangled channel in traditional RSP \mbox{protocols \cite{nguyen2011remote, PhysRevA.98.042329, Du2019}}. Assuming $|\alpha|=sin\,\theta$, $|\beta|=cos\,\theta (0 \le \theta \le \frac{\pi}{4})$, the success probability can be represented by $2(sin\,\theta)^2$. Figure \ref{Fig.1} illustrates the variation of the success probability as a function of $\theta$. Undoubtedly, the closer $\theta$ is to $\frac{\pi}{4}$, the higher the success probability is. Only when $\theta=\frac{\pi}{4}$, i.e., employing a maximally entangled channel, does the probability of success reach 1 according to traditional quantum communication protocols. As mentioned above, there is still some challenge in preparing the ideal maximal entanglement. It is a necessity to design a refreshed protocol in which RSP always succeeds even if the quantum channel is a non-maximally entangled state ($|\alpha| \neq |\beta|$).

section{Deterministic RSP via a Generally Entangled Quantum Channel}

To achieve a deterministic RSP via a non-maximally entangled channel, we have designed a fresh RSP scheme wherein a general quantum state is faithfully prepared in a remote place by using a non-maximal entanglement. The success probability of RSP is increased from $2|\alpha|^{2} (|\alpha|^{2}\leq\frac{1}{2})$ to 1 by means of an auxiliary qudit and simpler measurement approaches. The next sections are devoted to the preparation of 2-dimensional quantum states as an example to introduce the scheme proposed in this paper in detail and then to extend it to arbitrary dimensions.

\subsection{Deterministic RSP of a 2-Dimensional Quantum State via a Generally Entangled \mbox{Quantum Channel}}

Suppose Alice aims to prepare a two-dimensional quantum state for Bob,
\begin{eqnarray}
    \begin{aligned}
        |\varphi\rangle^{(2)}&=x_{0}|0\rangle+x_{1}|1\rangle\\
                       &=x_{0}|0\rangle+|x_{1}|e^{i\theta}|1\rangle, \label{equation 11} 
    \end{aligned}
\end{eqnarray}
where $x_{0}$ is a real number, $x_{1}$ is a complex number, and both satisfy the orthogonal normalization $x_{0}^{2}+|x_{1}|^{2}=1$. The employed quantum state in the protocol is the entangled quantum state of qubits A and B $|\varphi\rangle^{(2)}_{AB}=(\alpha|00\rangle+\beta|11\rangle)_{AB}$ ($|\alpha|^{2}+|\beta|^{2}=1$, $|\alpha|\leq|\beta|)$. In the beginning, Alice holds qubits A and B. The employed quantum state in the protocol is the entangled quantum state of qubits A and B:
\begin{eqnarray}
    \begin{aligned}
|\varphi\rangle^{(2)}_{AB}=(\alpha|00\rangle+\beta|11\rangle)_{AB}, 
\end{aligned}
\end{eqnarray}
where $\alpha$ and $\beta$ are complex numbers, $|\alpha|^{2}+|\beta|^{2}=1$ and $|\alpha|\leq|\beta|$. Only when \linebreak $|\alpha|=|\beta|$ is $|\varphi\rangle^{(2)}_{AB}$ a maximally entangled quantum state; otherwise, $|\varphi\rangle^{(2)}_{AB}$ is a non-maximally entangled quantum state. Additionally, Alive introduces an auxiliary qubit C set on $|0\rangle_{C}$ initially and performs the C-NOT gate $C^{(2)}_{AC}$ on qubits A and C in the same manner as the previous RSP schemes. 

Step (\uppercase\expandafter{\romannumeral1}) Because she knows the information prepared for Bob, Alice applies the operation $U^{(2)}_{A}=(x_{0}|0\rangle+|x_{1}|e^{i\theta}|1\rangle)\langle 0|+(-|x_{1}|e^{i\theta}|0\rangle+x_{0}|1\rangle)\langle 1|$ on qubit A. After that, the hyperentangled state unitarily evolves into
\begin{eqnarray}
 \begin{aligned} 
    |\varphi_1\rangle^{(2)}_{ABC}=&U^{(2)}_{A}|\varphi_0\rangle^{(2)}_{ABC}\\
                           =&(\alpha x_{0}|000\rangle+\alpha |x_{1}|e^{i\theta}|100\rangle\\
                             &-\beta |x_{1}|e^{i\theta}|011\rangle+\beta x_{0}|111\rangle)_{ABC}.   
 \end{aligned}
\end{eqnarray}

Note that $U^{(2)}_{A}$ is an element of the two-dimensional special unitary group $SU_{2}$ constituted by a set of 2 $\times$ 2 complex matrices, which have a determinant of unity. Any one of these elements can be generated by a set of generators \{$I,\sigma_{x},\sigma_{y},\sigma_{z}$\} \cite{PhysRevA.51.1015, PhysRevLett.74.4087, PhysRevA.52.3457}, i.e., $U^{(2)}_{A}$ can be substituted by these generators.

Step (\uppercase\expandafter{\romannumeral2}) Alice performs C-NOT gates $C^{(2)}_{AB}$ and $C^{(2)}_{BA}$ on qubits A and B.
\begin{eqnarray}
    \begin{aligned} 
        |\varphi_2\rangle^{(2)}_{ABC}=&C^{(2)}_{BA}C^{(2)}_{AB}|\varphi_1\rangle^{(2)}_{ABC}\\
        =&(\alpha x_{0}|000\rangle+\alpha |x_{1}|e^{i\theta}|010\rangle\\
        &-\beta |x_{1}|e^{i\theta}|111\rangle+\beta x_{0}|101\rangle)_{ABC}\\
        =&\alpha|00\rangle_{AC}(x_{0}|0\rangle+|x_{1}|e^{i\theta}|1\rangle)_{B}\\
        &+\beta |11\rangle_{AC}(x_{0}|0\rangle-|x_{1}|e^{i\theta}|1\rangle)_{B}.
\end{aligned} 
\end{eqnarray}

Step (\uppercase\expandafter{\romannumeral3}) Next is an essential and crucial step in the quantum communication schemes, i.e., entanglement distribution. To establish a quantum channel between the two communicators, Alice distributes qubit B to Bob. Then, an entanglement is established between these two communicators. Qubits A and C are still held by Alice.

Step (\uppercase\expandafter{\romannumeral4}) The final step is the measurement. The measurement basis vector employed in our scheme is different from the one used in the traditional RSP schemes. We choose the simplest basis vectors $\{|0\rangle, |1\rangle\}$ instead of the information-related basis vectors. Alice performs the projective measurement on qubits A and C one by one, then informs the measurement results to Bob. According to the measurement results of qubits A and C, Bob performs the corresponding operation on particle B. If the result is $|00\rangle_{AC}$, Bob does nothing to reconstruct the target state (Equation (\ref{equation 11})); if the result is $|11\rangle_{AC}$, Bob performs the Pauli operator $\sigma_{z}$ on qubit B to reconstruct Equation (\ref{equation 11}). Thus, the success probability of RSP is $\alpha^{2}+\beta^{2}=1$. Bob obtains the target state (Equation (\ref{equation 11})) with 100\% probability.

\subsection{Deterministic RSP of a d-Dimensional Quantum State via a Generally Entangled \mbox{Quantum Channel}}

Suppose that Alice's mission is to prepare a d-dimensional quantum state $|\varphi\rangle^{(d)}$ to Bob, 
\begin{eqnarray}
    \begin{aligned}
        |\varphi\rangle^{(d)}=\sum_{n=0}^{d-1}x_{n}|n\rangle
        =\sum_{n=0}^{d-1} |x_{n}|e^{i\theta_{n}}|n\rangle \label{equation 1},
    \end{aligned} 
\end{eqnarray}
where $x_{n}$ are complex numbers satisfying the normalization condition $\sum_{n=0}^{d-1}|x_{n}|^{2}=1$. Qubits A and B are entangled and possessed by Alice originally. Their entangled quantum state $|\varphi\rangle^{(d)}_{AB}$ is represented by the following form, 
\begin{eqnarray}
    \begin{aligned}
        |\varphi\rangle^{(d)}_{AB}=\sum_{m=0}^{d-1} \lambda_{m}|mm\rangle_{AB} \label{equation 2},
    \end{aligned} 
\end{eqnarray}
where complex numbers $\lambda_{mn}$ satisfy orthogonal normalization $\sum_{m,n=0}^{d-1}|\lambda_{mn}|^{2}=1$. Only when all $|\lambda_{mn}|$ are equal is $|\varphi\rangle^{(d)}_{AB}$ a maximally entangled quantum state; otherwise, it is a non-maximally entangled state. Alice introduces an auxiliary qubit set C in the initial state $|0\rangle_{C}$ and employs a C-NOT gate on qubits A and C, which entangles these three particles and renders two separable quantum states $|\varphi\rangle^{(d)}_{AB}$ and $|0\rangle_{C}$ into a hyperentangled \mbox{quantum state},
\begin{eqnarray}
    \begin{aligned}
        |\varphi_{0}\rangle^{(d)}_{ABC}=&C_{AC}^{(d)}(|\varphi\rangle^{(d)}_{AB}\otimes|0\rangle_{C})\\
        =&\sum_{m=0}^{d-1}\lambda_{m}|mmm\rangle_{ABC},
    \end{aligned} 
\end{eqnarray}

The mathematical form of a generalised C-NOT gate can be written as
\begin{eqnarray}
    \begin{aligned}
        C^{(d)}_{ij}=\sum_{i,j=1}^{d-1}|i, j \oplus k_{i}\rangle_{ij ij} \langle i, j|,
    \end{aligned} 
\end{eqnarray}
where qubit i is the control qubit, and qubit j is the target qubit. When the quantum state of qubit i is $|i\rangle_{i}$, the quantum state of qubit j is changed from $|j\rangle_{i}$ to $|j \oplus k_{i}\rangle_{j}$, where $k_{i}=0,1,\cdots,d-1$. We hereby assume that $j \oplus k_{m}=m$ when $j=0$.

Step (\uppercase\expandafter{\romannumeral1}) As Alice knows the information transferred to Bob, she is allowed to perform the operation associated with the information $U^{(d)}_{A}=\sum_{n=0}^{d-1} |x_{n}|e^{i\theta_{n}}|n\rangle \langle m|$ on qubit A. After that, the hyperentangled state unitarily evolves into
\begin{eqnarray}
    \begin{aligned}
        |\varphi_{1}\rangle^{(d)}_{ABC}=&U^{(d)}_{A}|\varphi_{0}\rangle^{(d)}_{ABC}\\
                                    =&\sum_{m,n=0}^{d-1} \lambda_{m}|x_{mn}|e^{i\theta_{mn}}|nmm\rangle_{ABC}.
    \end{aligned} 
\end{eqnarray}

Step (\uppercase\expandafter{\romannumeral2}) To reach the target further, Alice applies C-NOT gates $C^{(d)}_{AB}$ ($m \oplus k_{n}=n$ when $i=n$) and $C^{(d)}_{BA}$ ($n \oplus k_{n}=m$ when $i=n$) on qubits A and B. 
\begin{eqnarray}
    \begin{aligned}
        |\varphi_{2}\rangle^{(d)}_{ABC}=&C^{(d)}_{BA}C^{(d)}_{AB}|\varphi_{1}\rangle^{(d)}_{ABC}\\
                                 =&\sum_{m, n=0}^{d-1} \lambda_{m}|x_{mn}|e^{i\theta_{mn}}|mnm\rangle_{ABC}\\
                                 =&\sum_{m=0}^{d-1}\lambda_{m}|mm\rangle_{AC}(\sum_{n=0}^{d-1}|x_{mn}|e^{i\theta_{mn}}|n\rangle_{B}).\label{equation 14}
    \end{aligned} 
\end{eqnarray}

Step (\uppercase\expandafter{\romannumeral3}) Next is the entanglement distribution. To establish a quantum channel between two communicators, Alice distributes qubit B to Bob. Then, an entanglement is established between these two communicators.

Step (\uppercase\expandafter{\romannumeral4}) Alice implements the projective measurement on qubits A and C, then tells Bob the measurement results. Bob applies the corresponding operation on qubit B based on the measurement results of qubits A and C to reconstruct the target state (Equation (\ref{equation 1})) Alice prepares for him. From Equation (\ref{equation 14}), the success probability of RSP is $\sum_{m=0}^{d-1}|\lambda_{m}|^{2}=1$.

\begin{figure*}[htbp]
    \centering
    \includegraphics[width=0.8\paperwidth]{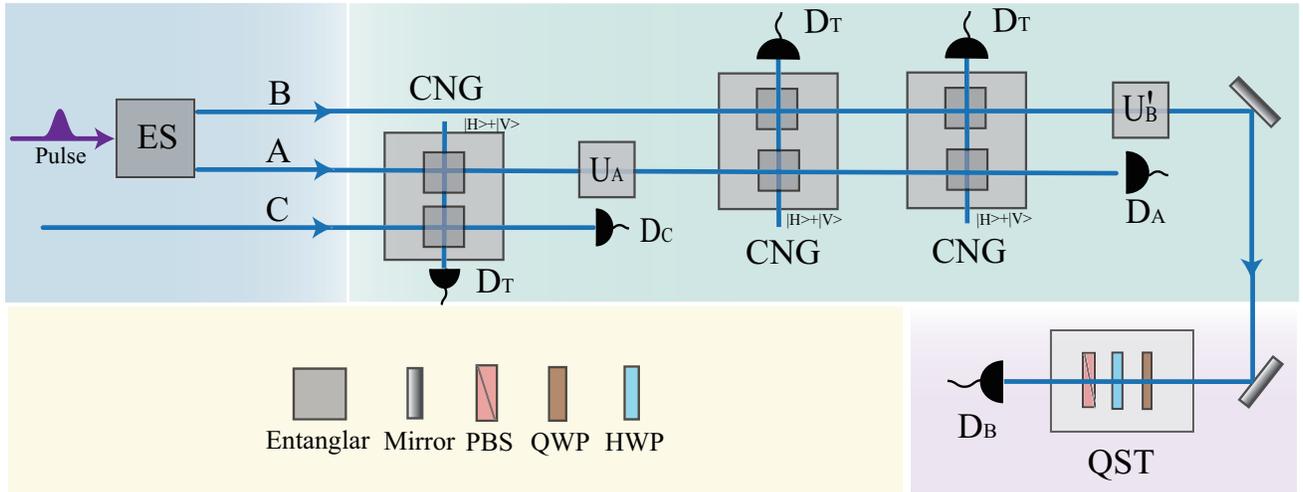}
    \caption{A schematic of deterministic RSP using a non-maximally entangled channel. The entanglement resource (ES) is employed to produce the polarization-entangled state of photons A and B. The polarization state of photon C is set on $|H\rangle_{C}$ initially. Alice first performs a C-NOT gate $C^{(d)}_{AC}$ on photons A and C. One extra ancilla photon $|H\rangle+|V\rangle$ is needed beyond the control and target photons to realize a deterministic C-NOT operation. Secondly, Alice applies a single-photon unitary transformation $U^{(d)}_{A}$ on photon A. Thirdly, she performs C-NOT gates $C^{(d)}_{AB}$ and $C^{(d)}_{BA}$ on photons A and B in succession. After that, Alice distributes photon B to Bob. Finally, the projective measurement is implemented on photons A and C, and the measurement results are sent to Bob. Bob performs the corresponding operations on photon B according to the measurement results. The quantum state of photon B is analyzed via QST to test the quality of this communication scheme. Note that CNG represents the C-NOT gate, PBS represents the polarizing beam splitter, QWP represents the quarter-wave plate, and HWP represents the half-wave plate.}\label{Fig.2}
\end{figure*}
\section{Realization}
The schematic diagram for transporting a polarization-encoded photon via a generally entangled channel is illustrated in Figure \ref{Fig.2}. Suppose the assignment of Alice is to facilitate Bob in preparing a polarization state of a photon: $|\varphi\rangle=x_{0}|H\rangle+x_{1}|V\rangle$ (where H indicates the horizontal linear polarization and V indicates the vertical linear polarization).  {The polarization state of photon C is set on $|H\rangle_{C}$ initially \cite{PhysRevA.67.040301, PhysRevLett.93.093601, PhysRevLett.100.133601, PhysRevLett.105.030407, Wang2015}. Here, we adopt the spontaneous parametric down-conversion (SPDC) process based on nonlinear crystals to prepare the polarization entangled state of photons A and B. SPDC is a nonlinear parametric process in which a strong pump light and nonlinear crystal interact to generate a signal and an idle photon \cite{PhysRevLett.25.84, PhysRevLett.75.4337, PhysRevA.60.R773, PhysRevA.50.R2803}. As the down-conversion photon owns the time, polarization, frequency, and spin entanglement, SPDC is often used to generate specific polarization entanglement states, which are the basis of the quantum information experiment and its applications \cite{Bouwmeester1997, Pan2000, Yin2012, Ren2017, PhysRevLett.117.170403, PhysRevLett.125.230501, Zhang2022}.} After the completed preparation of the polarized entangled state and a single-photon polarized quantum state, Alice performs a C-NOT gate on the control photon A and the target photon C. Note that one auxiliary photon set in the polarized state $|H\rangle+|V\rangle$ is necessary for the performance of a faithful CNOT operation \cite{PhysRevLett.93.250502, PhysRevLett.94.030501, yamamoto2003demonstration}. This ancilla is unconsumed in the gate operation and can be recycled for further use. Because Alice knows the prepared quantum information for Bob, she possesses the capability to apply the unitary operation $U^{(d)}_{A}$ associated with the information on photon A. This operation can be decomposed into a combination of Hadamard gates and phase gates \cite{PhysRevA.84.050301, PhysRevA.87.012307, Larsen, PhysRevLett.83.5166, PhysRevLett.106.013602, PhysRevLett.124.160501}. Alice applies C-NOT gates $C^{(2)}_{AB}$ and $C^{(2)}_{BA}$ on photons A and B. Next is the entanglement distribution and measurement.  The projective measurement is implemented on photons A and C, and the measurement results are sent to Bob. According to the results, Bob performs the corresponding operations $U_{B}^{(d)}$ on photon B to reconstruct the target state (Equation (\ref{equation 1})). The quality of RSP is estimated by performing quantum state tomography (QST) between the prepared polarized state $|\varphi\rangle=x_{0}|H\rangle+x_{1}|V\rangle$ and the obtained polarized state of photon B \cite{PhysRevLett.74.4101, PhysRevLett.105.150401, PhysRevLett.126.100402}.

\begin{figure}[htbp]
    \includegraphics[width=0.4\paperwidth]{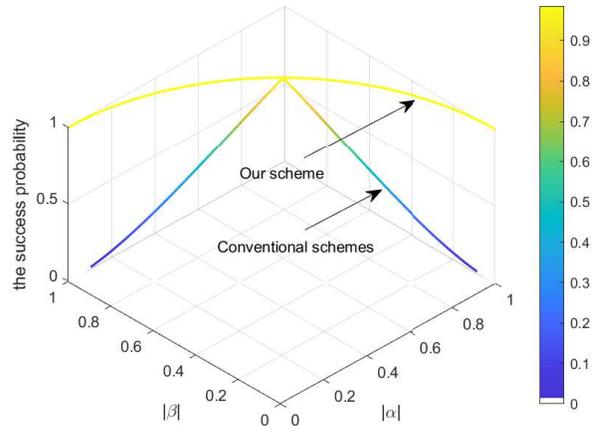}
    \caption{The success probability is a function of entanglement coefficients of the quantum channel. As the quantum channel degenerates from a maximally entangled state $(|\alpha|=|\beta|=\frac{1}{\sqrt{2}})$ to a partially entangled one $(|\alpha|\neq |\beta|)$, the success probability of conventional schemes is decayed accordingly, but the success probability of our perfect transport scheme remains 1 and constant.}\label{Fig.3}
\end{figure}

\section{Discussion}

Herein, we offer an alternative solution and demonstrate a fresh RSP protocol that allows for increasing the success probability of RSP from $2|\alpha|^{2} (|\alpha|^{2}\leq\frac{1}{2})$ to 100\% without additional consumption of quantum resources. The perfect transfer of information encoded in quantum states demands ideal maximally entangled channels. Most attempts to improve communication efficiency involve enhancing the entanglement strength of quantum channels by investing more quantum resources. This solution requires no prior improvement of the entanglement channel. In most conventional RSP schemes \cite{nguyen2011remote, PhysRevA.98.042329, Du2019}, the probability of success is affected by the coefficients of the entanglement channel. Only when the quantum channel is in a maximally entangled state, i.e., $|\alpha|=|\beta|$, can the probability of success reach 1, as illustrated in Figure \ref{Fig.3}. However, the success probability in this project is independent of the coefficients of the quantum channel. Even if the utilized quantum channel is degraded to a non-maximally entangled one, i.e., $|\alpha|\neq |\beta|$, the success probability of RSP remains 1 and is always constant without increasing quantum resource consumption, as displayed in Figure \ref{Fig.3}. Furthermore, this ingenious protocol only craves single communication, whereas Roa's scheme \cite{PhysRevA.91.012344} needs multiple correspondences to transfer an equivalent amount of quantum information. It is unnecessary to construct multiple and repeated entanglement channels for the realization of deterministic communication assignments \cite{PhysRevA.73.022340}. In addition, we adopted a more convenient measurement method of PM instead of the complex approach of positive operator-valued measurement \cite{PhysRevA.98.042329}. Our generalized scheme applies to preparing quantum states of arbitrary dimensions and is not limited to the preparation of 2, 4, and 8-dimensional quantum states \cite{PhysRevA.65.022316}. It is well-known that the ability of coherent manipulation of higher-dimensional quantum states is crucial for inventing advanced quantum technologies \cite{Dada2011, Erhard2018, Erhard2020}. Concerning the circumscribed two-dimensional systems, higher-dimensional systems possess virtues of noise resilience in quantum communications tasks and are more efficient than lower-dimensional systems in quantum computing and simulation. Consequently, this extension transmission of realizable high-dimensional quantum states is significant for developing quantum science and technology.

\section{Conclusions}

We have addressed imperfect RSP using a generally entangled channel and provided a noise-resistant faithful deterministic protocol. The sender transfers quantum information to the receiver deterministically even if the quantum channel is a non-maximally entangled quantum state. The success possibility of communication is boosted to 100\% without more consumption of quantum resources, which is beyond the standard protocol threshold. We are unconditional on heightening the entanglement of quantum channels as before. Moreover, our solution is more concise and realizable since we employ the generally entangled quantum channel and a simple measurement method concerning the previous protocols. Our approach has provided a deeper insight into interesting extensions of traditional schemes and boasts better performances in counteracting environmental noises in quantum communication tasks. This study presents brilliant thinking for transferring information more efficiently in practical quantum networks. Based on several advanced experimental attempts, we have devised schematics for preparing a two-dimensional quantum state, for example, transferring a polarization-encoded photon via a generally entangled channel. In short, we have designed a feasible communication scheme that can solve the problem of communication efficiency degradation with entanglement, which is constructive for the implementation of actual quantum networks. Further studies could fruitfully extend this to other quantum communication tasks and design more general solutions addressing the impact of noise channels on practical communication efficiency.

\bibliography{literature}

\end{document}